\documentclass[aps,twocolumn,epsfig,graphics,showpacs,floatfix]{revtex4}
\usepackage{amsmath,amsfonts,amssymb,graphics,graphicx,epsfig,color,times}

\begin{document}
\title{High-Temperature Atomic Superfluidity in Lattice Boson-Fermion Mixtures}
\author{Fabrizio Illuminati and Alexander Albus} 
\affiliation{
Dipartimento di Fisica, Universit\`{a} di Salerno, INFM - UdR di Salerno, and INFN, \\
Sezione di Napoli - Gruppo collegato di Salerno, Via S. Allende, I--84081 Baronissi
(SA), Italy}

\date{June 8, 2004}

\begin{abstract}
We consider atomic Bose-Fermi mixtures in optical lattices
and study the superfluidity of fermionic atoms
due to $s$-wave pairing
induced by boson-fermion interactions. 
We prove that the induced fermion-fermion coupling is always 
{\it attractive} if the boson-boson on site interaction is
repulsive, and predict the existence of an enhanced  BEC--BCS crossover 
as the strength of the lattice potential is varied.
We show that for direct on-site fermion-fermion {\it repulsion},
the induced attraction can give rise to superfluidity 
via $s$-wave pairing, at striking variance with the case of pure 
systems of fermionic atoms with direct repulsive interactions.
\end{abstract}


\maketitle

The recent spectacular acceleration in the experimental
manipulation \cite{Bloch1,Kasevich,Bloch2,Modugno}
and theoretical understanding \cite{Jaksch,Zwerger}
of systems of neutral atoms in optical lattices 
is leading to new and far reaching possibilities in
the study of complex systems of condensed matter
physics. Besides opening the way to the controlled
simulation and experimental testing of models of strongly correlated
systems \cite{Jaksch,Hofstetter}, such as, e.g., high-$T_{c}$ superconductors
and Hall systems, atomic physics in 
optical lattices hints at the possibility of 
discovering and probing new quantum phases of matter. 
Among others, very interesting recent studies in this 
direction are concerned with
lattice boson-fermion mixtures
\cite{Albus,Buechler,Lewenstein,Burnett,Cramer}.

Optical lattices are stable periodic arrays of 
microscopic potentials created by the interference
patterns of intersecting laser beams.
Atoms can be confined to different lattice sites,
and by varying the amplitude of the periodic potential,
it is possible to tune the
interatomic interactions at will. Therefore, optical
lattices provide an ideal tool to reach strong
coupling regimes even in the dilute limit \cite{Zwerger}.
The theory of neutral bosonic atoms in optical lattices has
been developed \cite{Jaksch} by assuming that the atoms are
confined to the lowest Wannier band of the periodic potential.
It can then be shown \cite{Jaksch} that the system is effectively described by a
single--band Bose--Hubbard model Hamiltonian \cite{Fisher}.
In such a model the superfluid--insulator transition is predicted
to occur when the on-site boson--boson interaction energy becomes
comparable to some multiple of the hopping energy between 
adjacent lattice sites.
This situation can be experimentally achieved by increasing the strength
of the lattice potential, which results in a strong suppression of the
kinetic (hopping) energy term.
In this way, the superfluid--Mott-insulator quantum phase
transition has been realized in a series of beautiful experiments
by loading an ultracold atomic Bose--Einstein condensate in an optical
lattice \cite{Bloch2}.
The theory of dilute mixtures of interacting bosonic and fermionic neutral
atoms subject to an optical lattice has been recently developed
by deriving an effective single--band Bose--Fermi Hubbard (BFH)
Hamiltonian from the underlying microscopic many-body dynamics \cite{Albus}. 
The zero-temperature phase diagram
of the BFH model has been analyzed
both in the homogeneous and inhomogeneous cases
\cite{Albus,Buechler,Lewenstein,Burnett,Cramer}.
These recent studies have shown that Bose-Fermi mixtures
are very fundamental systems of condensed matter exhibiting
complex phase diagrams and a potentially very rich physics
that is just beginning to be unravelled.

The quest for superfluidity of fermionic atoms in dilute, degenerate
quantum gases is the current object of widespread theoretical and experimental
efforts \cite{Hofstetter,Stringari,Thomas}. 
In the present work we propose
a new mechanism for fermion pairing in atomic systems.
We consider mixtures of bosonic and fermionic
atoms in optical lattices. The fermions are assumed to belong
to the same atomic species, but trapped in different spin states,
so that they can differ by angular momentum or generalized 
spin $\sigma$ ($\sigma =\{\uparrow,\downarrow\}$).
This situation can be easily achieved, for instance, by all--optical 
trapping. 
The mixture is then characterized by independent bosonic and
fermionic hoppings and by direct on-site boson-boson, 
boson-fermion, and fermion-fermion couplings.
For such a system we determine the structure of the
induced fermion-fermion coupling due to the presence
of the bosons, in analogy with recent and important studies on
the effects of induced interactions in boson-fermion mixtures
in free space \cite{Viveritall}.
We show that the regime of static
linear response is an excellent approximation when the bosons
are fully superfluid (weak optical lattice potential for
the bosons) and, provided that the 
bosons are endowed with repulsive
self-interactions, we find that the induced on-site 
fermion-fermion interaction
is always attractive, thus leading to local pairing,
irrespective of the sign of the boson-fermion interaction.
In the case of a direct, {\it repulsive} on-site fermion-fermion
interaction we show that the addition of the induced,
attractive fermion-fermion interaction leads to
an enhanced crossover regime between extended BCS pairing
and Bose-Einstein condensation of local molecules, and allows
for the occurrence of $s$-wave superfluidity of fermionic
atoms. This is at striking variance with the case of pure Fermi systems
with {\it repulsive} interactions, for which $s$-wave pairing
is impossible and superfluidity might be achieved, at much lower
temperatures, via higher-order mechanisms such as anisotropic
$d$-wave pairing.

The single-band BFH system in an optical lattice
with nearest-neighbor hopping and on-site direct boson-boson,
boson-fermion, and fermion-fermion interaction
is described by the Hamiltonian \cite{Albus}
\begin{eqnarray}
\label{Hubbard}
\hat{H} & = & -J_{B}\sum_{\{i,j\}}
\left( \hat{a}_{i}^{\dagger}\hat{a}_{j} + \mbox{h.c.} \right)
-J_{F}\sum_{\{i,j\},\sigma}
\left( \hat{c}_{i,\sigma}^{\dagger}\hat{c}_{j,\sigma} + \mbox{h.c.} \right)
\nonumber \\
& + & \frac{U_{BB}}{2}
\sum_i\hat{n}_{i}^{B}(\hat{n}_{i}^{B} - 1)
+ U_{BF}\sum_i\hat{n}_{i}^{B}\hat{n}_{i}^{F}
\nonumber \\
& + & U_{FF}^{dir}\sum_{i}\hat{n}_{i,\uparrow}\hat{n}_{i,\downarrow} \; ,
\end{eqnarray}
where $J_{B}$ and $J_{F}$ are, respectively the boson
and fermion hopping amplitudes between adjacent lattice sites, and $U_{BB}$, $U_{BF}$, and
$U_{FF}^{dir}$ are, respectively the direct boson-boson, boson-fermion,
and fermion-fermion on-site interaction strengths.
The operators $\hat{a}_{i}$ are the on-site bosonic annihilation
operators, and $\hat{c}_{i,\sigma}$ are the on-site fermionic
annihilation operators for states with spin $\sigma$. The bosonic
on-site occupation number is $\hat{n}_{i}^{B} = 
\hat{a}_{i}^{\dagger}\hat{a}_{i}$, while $\hat{n}_{i,\sigma}
= \hat{c}_{i,\sigma}^{\dagger}\hat{c}_{i,\sigma}$ denotes
the fermionic on-site occupation number for states 
with spin $\sigma$, and, finally, $\hat{n}_{i}^{F}
= \hat{n}_{i,\uparrow} +\hat{n}_{i,\downarrow}$ denotes
the total fermionic on-site occupation number.
The bosonic and fermionic lattice potentials $V_{B}$ and $V_{F}$
are produced by a standing laser wave: 
$V_{B,F}(x,y,z)=
V_{B,F}(\sin^{2}kx + \sin^{2}ky + \sin^{2}kz)$, 
where $k=\pi/a$ is the wave vector of the laser
field, $a$ denotes the lattice spacing, and the 
lattice stregths $V_{B,F}$ that are felt by the bosons
and the fermions depend on the intensity of
the laser light and the detuning 
of the laser fequency against the respective atomic frequencies.
Scaling the tunable lattice
amplitudes $V_{B,F}$ with the bosonic and fermionic
recoil energies, respectively 
$E^R_{B} = (\pi\hbar)^{2}/2a^{2}m_{B}$ and
$E^R_{F} = (\pi\hbar)^{2}/2a^{2}m_{F}$, 
one can define the dimensionless bosonic
and fermionic lattice amplitudes $\eta_{B,F}
=V_{B,F}/E^R_{B,F}$. Given the wavelength $\lambda_L$ of
the laser field, and the internal-state wavelengths 
$\lambda_B$, $\lambda_F$ for bosons and fermions, one
has $\eta_F/\eta_B = ((\lambda_F/\lambda_B)^{4}\cdot
(\Gamma_F \Delta \lambda_B E^{R}_{B}))/(\Gamma_B \Delta
\lambda_F E^{R}_{F})$, where $\Delta \lambda_{B,F} = 
\lambda_{B,F} - \lambda_{L}$ are the boson and fermion
detunings from the laser field, and the ratio of the
natural line widths $\Gamma_F/\Gamma_B$ is usually of 
order one. By properly adjusting the detunings (blue or
red) with the gauging of the laser field wavelength $\lambda_{L}$,
one can realize lattices whose depths can either be very different
or nearly equal for the bosons and the fermions.
The hopping and on site energy amplitudes  
exhibit a characteristic behavior as
a function of the parameters of the optical
lattice: $J_{B,F} = E_{B,F}^{R}(2/\sqrt{\pi})
\eta_{B,F}^{3/4}\exp{(-2\eta_{B,F}^{1/2})}\,$;
$U_{BB,FF} = E_{B,F}^{R}a_{BB,FF}k\sqrt{8/\pi}
\eta_{B,F}^{3/4}\,$, where $a_{BB}$ and $a_{FF}$
are the direct boson-boson and fermion-fermion
$s$-wave scattering lengths, and
$U_{BF} = E_{F}^{R}a_{BF}k(4/\sqrt{\pi})(1+m_F/m_B)
(\eta_F^{-1/2}+\eta_B^{-1/2})^{-3/2}\,$, 
where $a_{BF}$ is the direct boson-fermion $s$-wave 
scattering length. The ratio of the hopping amplitudes
reads $J_B/J_F =  [(E_{B}^{R}\eta_B^{3/4})/(E_F^R\eta_F^{3/4})]
\cdot \exp{[-2(\sqrt{\eta_B} - \sqrt{\eta_F})]}$.

The bosonic and fermionic dispersion relations in a
$3$-D optical lattice with nearest-neighbor hopping take
the following form:
\begin{eqnarray}
\epsilon_{B,F}(\mbox{\bf q}) & = &
4J_{B,F} \left[ \sin^{2}{\frac{q_{x}a}{2}} + \sin^{2}{\frac{q_{y}a}{2}}  
+ \sin^{2}{\frac{q_{z}a}{2}} \right] \; , 
\end{eqnarray}
where $J_{B}$ and $J_{F}$ are, respectively,
the bosonic and fermionic hopping amplitudes.
The bosonic energy spectrum in the Bogoliubov 
approximation reads
\begin{equation}
\omega_{B}(\mbox{\bf q}) = 
\sqrt{\epsilon_{B}(\mbox{\bf q})\left[
\epsilon_{B}(\mbox{\bf q}) + 2n_{B}U_{BB} 
\right] } \; .
\end{equation}
In the low momentum limit for the bosons, we recover
the standard quadratic single particle spectrum
$\epsilon_{B}(\mbox{\bf q}) \approx
3 J_{B}a^{2}q^{2}$,
where $q^{2} \equiv |\mbox{\bf q}|^{2}
= q_{x}^{2} + q_{y}^{2} + q_{z}^{2}$.
Correspondingly, the phononic Bogoliubov excitation 
spectrum in the long wavelength regime reads
$\omega_{B}(\mbox{\bf q}) \approx 
\sqrt{3 J_{B}(aq)^{2}2n_{B}U_{BB}}
= c_{B}q$, where $c_{B}=\sqrt{6 n_{B}J_{B}U_{BB}}a$ 
is the velocity of sound.

Let us next consider
the momentum and frequency dependent
bosonic dynamic response function
$\chi_{B}(\omega, \mbox{\bf q})$:
\begin{equation}
\chi_{B}(\omega, \mbox{\bf q}) = 
\frac{2n_{B}\epsilon_{B}(q)}{\omega^{2} 
- \omega_{B}^{2}(q)} \; .
\end{equation}
Regarding the effects induced on the fermions,
it is possible to show that the bosonic dynamic response function
can be replaced by its static limit approximation.
Let us focus on the fermion-fermion interaction
$U_{FF}^{ind}(\omega, \mbox{\bf q})$ induced by the bosons. 
If the frequency contribution
is not neglected, the gap function $\Delta(\omega, \mbox{\bf q})$
for the fermions depends on frequency averages over the
off-diagonal fermionic propagator $G_{12}(\omega, \mbox{\bf q})$.
\begin{equation}
\Delta(\omega, \mbox{\bf q})=\int U_{FF}^{ind}(\omega'-\omega, \mbox{\bf q}'-\mbox{\bf q}) 
     G_{12}(\omega', \mbox{\bf q}')
d \omega d \mbox{\bf q}'\; .
\end{equation}
The maximum scale of integration for the fermions
is fixed by the Fermi energy: 
$\hbar\omega_{max} \approx E_{F}$. 
The condition $\omega^{2}_{max} \ll 
\omega^{2}_{B}(q)|_{q=q_{F}}$ can be satisfied for
widely different scenarios of fermion filling 
factors.
We will consider 
half-band fermion filling $n_F \approx 1$.
This amounts to $q_{F}a \approx \pi/2$ and $E_{F} \approx 6J_{F}$.
We then have $\hbar^2\omega^{2}_{max} \approx 36 J_{F}^{2}$, and
$\omega^{2}_{B}(q)|_{q=q_{F}} \approx 36 J_{B}^{2} + 12J_{B}n_{B}U_{BB}$,
so that the static limit condition reads
\begin{equation}
3J_{F}^{2} \ll 3J_{B}^{2} + J_{B}n_{B}U_{BB} \; .
\label{inequality}
\end{equation}
This condition can be satisfied in several ways, depending on
the type of boson-fermion mixture that one actually considers,
and on the lattice strenghts and hopping amplitude ratios
that can be realized in experiments. For boson-fermion
mixtures with almost equal atomic masses, like ${}^{6}$Li-${}^{7}$Li,
and large red detuning of the laser field from the typical $D$ lines
of the Lithium isotopes, one has $J_{F} \simeq J_{B}$, which implies
a static limit condition of the form $6(J_{F}-J_{B})\ll n_{B}U_{BB}$. 
This condition is well satisfied at the onset of bosonic superfluidity
$U_{BB} \simeq 10J_{B}$ and bosonic filling factors $n_B \geq 1$.
Alternatively, the static limit condition can be satisfied by realizing 
small detunings such that $\eta_F > 2\eta_B$; this yields in general 
$J_B \gg J_F$, which again allows inequality (\ref{inequality}) to hold. 
In this second
instance, the bosonic lattice is usually some times weaker than the
fermion lattice, and the bosons are in the deep superfluid regime. This
setting is suited to treat mixtures irrespective of the atomic mass differences,
and is thus the one we will consider in the following when studying the degenerate
mixture of $^{40}$K and $^{87}$Rb.   
We will study 
quantitatively the case of half-band fermion filling
$n_{F} \approx 1$, because this is a situation that
can be easily realized experimentally, it provides the
optimal scenario for the highest attainable critical
temperatures, and finally allows a direct comparison with
the work of Hofstetter {\it et al.} on the high-temperature
superfluidity of pure fermionic systems in optical lattices
\cite{Hofstetter}. 
Taking the static bosonic response function 
$\chi_{B}(0, \mbox{\bf q}) = 
- 2n_{B}\epsilon_{B}(q)/\omega_{B}^{2}(q)$, the
resulting induced fermion-fermion interaction (due to 
the bosonic density-density fluctuations caused
by the boson-fermion coupling $U_{BF}$) reads 
$U_{FF}^{ind}(0,\mbox{\bf q}) = 
\chi_{B}(0, \mbox{\bf q}) U_{BF}^{2}$.
To discuss $s$-wave fermion pairing and evaluate the critical
temperature, one must compute the average induced interaction
${\bar{U}}_{FF}^{ind}$ over the Fermi surface \cite{Viveritall}. 
Provided that the bosonic lattice is not extremely weak 
({\it i.e.} $\eta_B$ not $\ll 1$), contributions from 
nearest-neighbor induced interactions are negligible in 
first approximation, and one finds
${\bar{U}}_{FF}^{ind} = 
- U_{BF}^{2}/U_{BB}$.
We see that, irrespective of the sign
of the boson-fermion $s$-wave scattering length $a_{BF}$, and thus
of the boson-fermion on-site coupling $U_{BF}$, the induced
fermion-fermion interaction is always {\it attractive} as
long as the bosons are endowed with a positive $s$-wave
scattering length $a_{BB} > 0$ and thus with a repulsive on-site
interaction $U_{BB} > 0$. This is the case we will
always consider in the following.
For distinguishable fermions, such as ensembles of
spin unpolarized neutral atoms, the total on-site 
fermion-fermion coupling $U_{FF}^{tot}$ is then  
\begin{equation}
U_{FF}^{tot} \equiv U_{FF}^{dir} + {\bar{U}}_{FF}^{ind} = U_{FF}^{dir}   
- \frac{U_{BF}^{2}}{U_{BB}} \; .
\label{Utot}
\end{equation}
Thus, regarding interatomic pairing, 
fermions in a lattice Bose-Fermi mixture  
behave as interacting particles
in a pure lattice fermionic system, but for the
crucial difference of acquiring a dressed attraction
that modifies the direct fermion-fermion interaction
through the boson-fermion coupling. This has relevant
consequences on the possible pairing mechanisms in the
system, in view of the
relative sign between the direct and the induced part
of the coupling strength. Because the sign of $U_{FF}^{dir}$
is determined solely by the sign of the fermion-fermion $s$-wave
scattering length $a_{FF}$, we need to treat separately 
the two possible instances 
$a_{FF} < 0$ of direct on-site attraction, and $a_{FF} > 0$
of direct on-site repulsion.

\begin{figure}
\setlength{\unitlength}{1cm}

\begin{picture}(8.5,5.5)(0,0)

\put(0.0,0.1){
  \epsfxsize=8.00cm
  \epsfbox{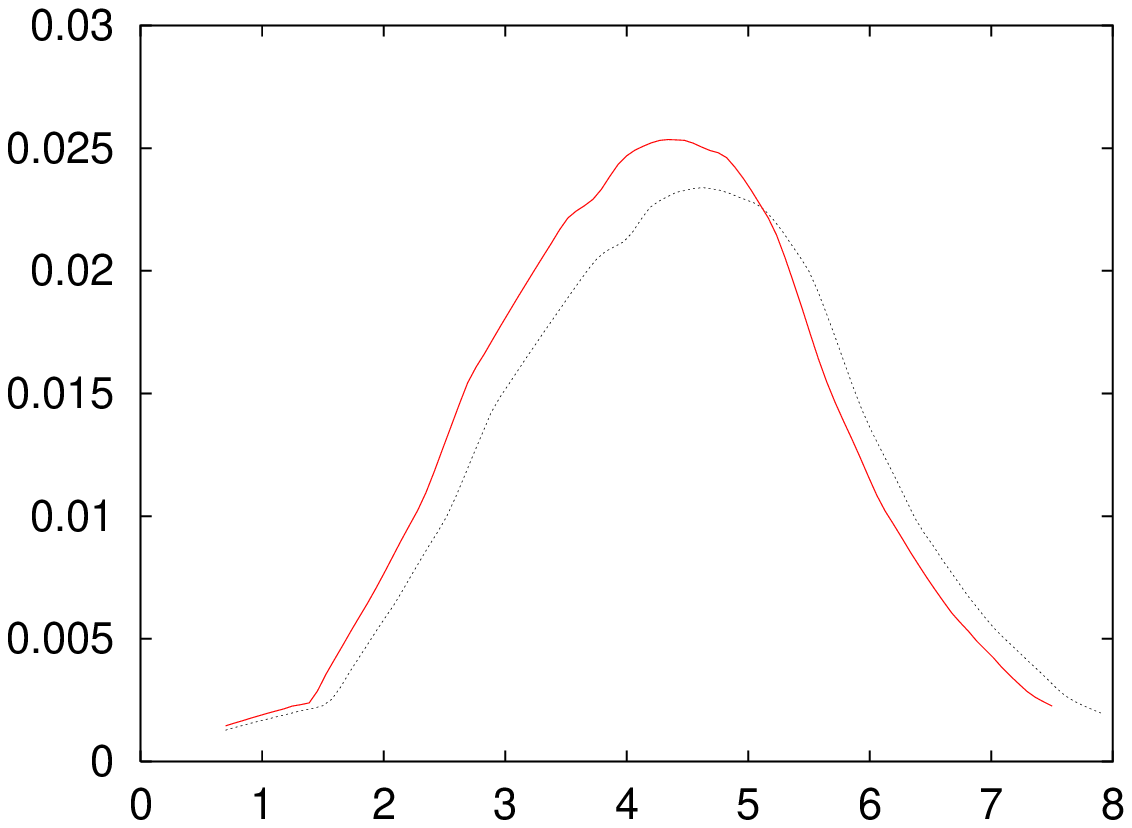}}
\put(0.0,3.0){\rotatebox[origin=c]{90}{
\makebox(0,0){$T_c [E_F^R/k_B]$}}}
\put(4.5,0.0){\makebox(0,0){$V_F/E_F^R$}}
\end{picture}
\caption
{Critical temperature for the SF transition of
$^6$Li atoms as a function of the fermion lattice strength
(dotted line), and of $^6$Li mixed with bosonic $^7$Li
(continuous line).}
\label{figure1}
\end{figure}

I) {\it Fermions with direct on-site attraction} --
In this case the fermions can undergo a transition  
to $s$-wave superfluidity at a higher critical
temperature $T_{c}$ than the analogous
pure Fermi case, as the direct and induced
interactions are both attractive and add together
(See Eq.~(\ref{Utot})). All the
considerations valid for pure fermionic systems
\cite{Hofstetter,Ranninger}
hold as well in the case of a Bose-Fermi mixture,
with the crucial difference that the
thermodynamic properties will now depend on the
total fermion-fermion interaction $U_{FF}^{tot}$ rather than the
direct one $U_{FF}^{dir}$.
Starting with a lattice of low or intermediate
depth, the BCS picture for $s$-wave pairing holds, predicting a critical
temperature for the transition with a scaling
behavior $k_{B}T_{c} \simeq 6J_{F}\exp{(-7J_{F}/|U_{FF}^{tot}|)}$. 
As the depth of the lattice is increased, 
the on-site interaction becomes at first comparable and then
finally dominates over the tunneling amplitude; then
the fermionic atoms form localized on-site bosonic
molecules that can undergo a Bose-Einstein condensation
(BEC) into the superfluid state at a transition temperature 
$k_{B}T_{c} \simeq 6J_{F}^{2}/|U_{FF}^{tot}|$.
In this situation the reduced ability of the pairs to
move around in the lattice leads to a net decrease of
the critical temperature. The maximum of the critical temperature
$T_{c}^{max}$ is then achieved at the region of crossover 
between the  BCS and BEC regimes,
$|U_{FF}^{tot}| \simeq 10J_{F}$ \cite{Scalettar,Ranninger}.
In Fig.~\ref{figure1} we show the behavior of the critical
temperature for the superfluid (SF) transition of $^6$Li atoms
as a function of the strength of the optical lattice, at
$n_{F} \simeq 1$.
When trapped in the state $|\uparrow, (\downarrow)\rangle$, 
the fermionic Lithium atoms are endowed with a very large and 
negative scattering length $a_{FF} \simeq -2500a_{0}$. 
This is the most favorable case of
direct fermion-fermion attraction known so far, 
and considered by Hofstetter {\it et al.} \cite{Hofstetter}.
Their result is compared in Fig.~\ref{figure1} with the
case of a $^6$Li-$^7$Li mixture, for which 
$a_{BF} \simeq 38a_{0}$ \cite{Schreck}. We see that the presence of
the bosons further enhances the maximum critical temperature
attainable for the SF transition of the fermionic atoms.
For a lattice spacing $a=10^{4}a_{0}$ the maximum critical temperature
$T_{c}^{max} = 0.11 \mu K$ for pure $^6$Li; for the mixture $^6$Li-$^7$Li 
we have $T_{c}^{max} = 0.14 \mu K$.

\begin{figure}
\setlength{\unitlength}{1cm}
\begin{picture}(8.5,5.5)(0,0)
\put(0.0,0.1){
  \epsfxsize=8.00cm
  \epsfbox{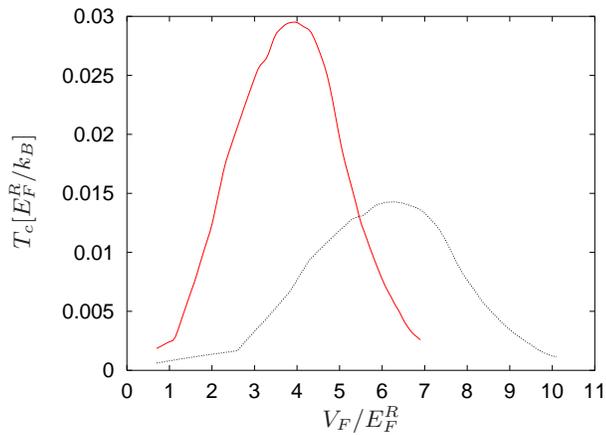}}
\put(0.0,3.0){\rotatebox[origin=c]{90}{
\makebox(0,0){$T_c [E_F^R/k_B]$}}}
\put(4.5,0.0){\makebox(0,0){$V_F/E_F^R$}}
\end{picture}
\caption{Critical temperature for the SF transition of
         $^{40}$K atoms mixed with $^{87}$Rb as a function 
         of the fermion lattice strength without a Feshbach resonance 
         (dotted line) and with a Feshbach resonance at $725$ G 
         (continuous line).}
\label{figure2}
\end{figure}

II) {\it Fermions with direct on-site repulsion} --
In the case of direct on-site repulsive fermion-fermion
interactions, $U_{FF}^{dir} > 0$, the picture departs
more radically from the analogous pure Fermi system.
The direct repulsion and the induced attraction
that determine the total interaction strength are now in
competition, and the net total sign depends on their relative
importance. In fact, because $U_{FF}^{dir}$ and $U_{BB}$ are both
positive, we can evaluate the total fermion-fermion 
interaction as $U_{FF}^{tot} =
(U_{BB} U_{FF}^{dir} - U_{BF}^{2})/U_{BB}$.
In several important instances of mixtures of alkali atoms
of current experimental interest, with direct fermion-fermion
repulsion, the measured value of the boson-fermion $s$-wave
scattering length $a_{BF}$ turns out to be much larger 
than both the boson-boson and fermion-fermion $s$-wave scattering lengths
$a_{BB}$ and $a_{FF}$, with the largest difference 
attained in degenerate mixtures of fermionic ${}^{40}$K
and bosonic ${}^{87}$Rb. Namely, in this case one has that 
$a_{FF} = 104.8 \pm 0.4a_{0}$ ($a_{0}$ being the Bohr radius) in
the singlet state, and $a_{FF} = 174 \pm 7a_{0}$ in the triplet
state \cite{Jin}, while $a_{BB} = 100.2 \pm 1.8a_{0}$, and 
$a_{BF} = -410 \pm 80a_{0}$ \cite{Modugno2}. 
A similar situation occurs for a mixture of 
fermionic ${}^{6}$Li with direct repulsion
and bosonic ${}^{7}$Li, where one has 
$a_{BB} \simeq a_{FF} \simeq 5a_{0}$, and $a_{BF} \simeq 38a_{0}$.  
In these important instances the direct  
fermion-fermion repulsion is thus dominated by the induced
fermion-fermion attraction, 
and high-temperature superfluidity of the fermionic atoms 
occurs by $s$-wave pairing.
This is at striking variance
with the case of systems of pure fermionic atoms with direct
on-site repulsion: for such systems
$s$-wave pairing is obviously always forbidden, and superfluidity 
can be achieved via different mechanisms of higher order,
at lower critical temperatures, such as anisotropic $d$-wave pairing.
Therefore, mechanism II) realizes the first atomic analogue of electron
paring in crystal lattices.
A further dramatic increase in the transition
temperature $T_{c}$ can be obtained by combining the boson-induced
interaction with the experimental manipulation of
the $s$-wave boson-fermion scattering length via Feshbach resonances.
In fact, the existence has been recently predicted \cite{Ferlaino}
of a Feshbach resonance at an applied magnetic field
$B \simeq 725$ G in the degenerate mixture of ${}^{40}$K and
${}^{87}$Rb, leading to an enhancement of the
boson-fermion $s$-wave scattering length to the very high
value $a_{BF} \simeq -687a_{0}$.
In Fig.~\ref{figure2} we show the behavior of the critical
temperature for the superfluid transition of $^{40}$K atoms
induced by the presence of the $^{87}$Rb atoms. Assuming the
isotopes trapped at the $D$-line values of current experiments
($795$nm for rubidium, $767$nm for potassium), and a blue-detuned
laser at $764$nm, we have $\eta_F \simeq 4 \eta_B$, so that
in the range of values of $\eta_F \simeq 7$ around the maximum 
critical temperature, we have $\eta_B \simeq 2$ (fully superfluid
bosons) and $J_B^{2} \simeq 20 J_F^{2}$ (static regime holds). 
We compare results for the case with (continuous curve) 
and without the presence of a Feshbach resonance (dotted curve).
For a lattice spacing $a=10^{4}a_{0}$ the first case yields
$T_{c}^{max} = 0.05 \mu K$, while without resonance we have 
$T_{c}^{max} = 0.04 \mu K$.
In conclusion, we have shown that for mixtures of bosonic and
fermionic neutral atoms in optical lattices, the fermionic
atoms acquire an induced on-site attraction mediated by the
boson-fermion interaction, irrespective of the sign of the latter.
This fact allows the fermions to undergo a 
high-temperature superfluid transition via total $s$-wave
pairing in both instances of direct attractive {\it and}
repulsive on-site fermion-fermion interactions.
Recent progresses in the manipulation of mixtures of bosonic
and fermionic atoms in optical lattices \cite{Modugno}
indicate that these predictions may be subject
to experimental verification in the near future.
We warmly thank Dr. K. Bongs for providing us with up-to-date
values on forthcoming experiments with degenerate mixtures
of $^{40}$K and $^{87}$Rb.
This work has been supported by the ESF under project
BEC2000+, the INFM, the INFN,
and the Italian Ministery for Scientific Research,
under project PRIN-COFIN 2002. 
We dedicate this work to Noam Chomsky on his 
75th birthday, for his tireless struggle in favor of free research,
human dignity and justice throughout the world.

\end{document}